\documentclass[superscriptaddress,11pt]{article}
\usepackage{authblk}
\usepackage{array}
\usepackage{type1cm}
\usepackage{lettrine}
\usepackage{graphicx}
\usepackage{geometry}
\usepackage{psfrag}
\usepackage{amsmath}
\usepackage{amsfonts}
\usepackage{appendix}
\usepackage[margin=10pt,font=footnotesize,labelfont=sc,format=hang,labelformat=parens]%
{caption}
\usepackage{subcaption}
\usepackage{amssymb}%
\providecommand{\U}[1]{\protect\rule{.1in}{.1in}}
\numberwithin{equation}{section}
\def\be{\begin{equation}}
\def\ee{\end{equation}}
\def\ba{\begin{eqnarray}}
\def\ea{\end{eqnarray}}
\def\bi{\begin{itemize}}
\def\ei{\end{itemize}}

\def\bra{\langle}
\def\ket{\rangle}

\def\E{\tilde E}
\def\Dp{              {}^{(+)}D         }
\def\Fp{       {}^{(+)}F                      }
\def\Ap{       {}^{(+)}A                  }
\def\Gp{              {}^{(+)}G        }
\def\Vp{       {}^{(+)}V                     }
\def\hp{       {}^{(+)}h                  }
\def\Dm{              {}^{(-)}D         }
\def\Fm{       {}^{(-)}F                      }
\def\Am{       {}^{(-)}A                  }
\def\Gm{              {}^{(-)}G         }
\def\Vm{       {}^{(-)}V                      }
\def\hm{       {}^{(-)}h                 }

\begin{document}

\title{From Euclidean to Lorentzian Loop Quantum Gravity via a Positive Complexifier}

\author{Madhavan Varadarajan}
\affil{Raman Research Institute\\Bangalore-560 080, India}

\maketitle

\begin{abstract}
We construct a positive complexifier, differentiable almost everywhere  on the classical phase space of real triads and $SU(2)$ connections, which generates a Wick Transform
from Euclidean to Lorentzian gravity everywhere except on  a phase space set of measure zero. 
This Wick  transform assigns an equal  role to the self dual and anti-self dual Ashtekar variables in quantum theory.
We argue that the appropriate quantum arena  for an analysis of the properties of the 
Wick rotation 
is the diffeomorphism invariant Hilbert space of Loop Quantum Gravity (LQG) rather than its kinematic Hilbert space.
We examine issues related to the construction, in quantum  theory, of the positive complexifier as a positive operator on
this diffeomorphism invariant Hilbert space.
Assuming the existence of such an operator, we explore the possibility of identifying 
physical states in Lorentzian LQG as Wick rotated images of physical states in the Euclidean theory. 
Our considerations derive from 
Thiemann's remarkable proposal to define Lorentzian LQG from  Euclidean LQG via the implementation in quantum theory of
a phase space `Wick rotation' which maps real Ashtekar-Barbero variables to Ashtekar's complex, self dual variables.

\end{abstract}

\section{\label{sec1}Introduction}
The Hamiltonian dynamics of classical General Relativity is generated by constraints i.e.  by functions on the gravitational phase space
which vanish on-shell. Canonical quantization seeks to implement the constraints as operators and identify their kernel with the space of physical states. 
The  complicated nonpolynomial form of the constraints in terms of the traditional  ADM phase space variables \cite{adm} provides an impediment to 
the development of a rigorous canonical quantization based on these variables. In contrast, the Ashtekar variables,  comprising of a complex connection and a conjugate triad 
render these constraints polynomial. Their discovery \cite{aanv} inspired a renewed attempt at  canonical quantization of gravity \cite{tedlee, leecarlo} leading to the development of 
the Loop Quantum Gravity (LQG) approach \cite{aabook,aajurekreview,ttbook,gpbook,apbook}. 

However, it has proven difficult to construct quantum representations based directly on the Ashtekar variables due to their
 complex valued nature.
As a result most developments in LQG are based on the related {\em real} Ashtekar-Barbero variables \cite{fer}. These variables are naturally adapted to Euclidean gravity and render its constraints
polynomial, this polynomiality being the analog of the polynomiality of the  constraints of the Lorentzian theory in the  complex  Ashtekar variables. 
In contrast, the expressions for the constraints of the {\em Lorentzian} theory  in terms of the {\em real} variables are  {\em non-polynomial}. However, due to the  impressive development of the quantum kinematics based on the real variables
there are concrete proposals which confront this non-polynomiality\cite{qsd,mastercon,jurekR}. 
%
Despite these advances in the construction
of the constraint operators for the Lorentzian theory based on their classical description in the  {\em real} Ashtekar-Barbero variables,  we are of the opinion that 
it would be preferrable to transit to a formalism which ascribes a central role to
the  original {\em complex} Ashtekar  variables. The reason is that the  complex Ashtekar variables have an immediate  spacetime interpretation: the  connection is the spatial pull back of the self dual part of the spacetime spin connection 
and the (densitized)
triad is  naturally related to the spacetime tetrad. In constrast the real Ashtekar-Barbero connection  has a purely spatial character with no direct spacetime interpretation \cite{sam}.
Hence it is natural to anticipate that the construction of a {\em spacetime covariant} quantum theory for Lorentzian gravity based entirely on the real variables would be significantly more complicated than  a construction
which incorporates the power and elegance of the self dual description.

In a remarkable set of papers in the mid-nineties \cite{qsd,ttwick,tttransform,ttlgrav}, Thiemann proposed a formalism which combines 
the rigor of the  quantum framework based on the real variables  with the simplicity of the classical description in terms of the complex variables.
He noted that the complex canonical transformation from real to complex
variables is generated by a {\em complexifier} function. This complexifier is  proportional to the spatial integral of the trace of the extrinsic curvature, the proportionality constant being imaginary
and equal to $-i \frac{\pi}{2G}$, $G$ being Newton's constant. The finite complex  canonical transformation generated by this complexifier  maps  the  real connection to its self dual counterpart. 
Since the Lorentzian constraints are obtained by substituting the real connection and triad in the Euclidean expressions by their images  through this complex canonical transformation, it follows that the 
complex canonical transformation maps the Euclidean constraints to their Lorentzian counterparts.  Thiemann also noted that the complexifier could be expressed as the Poisson bracket between the 
total volume $V$  of the spatial  slice and the Euclidean Hamiltonian constraint, $H_E(N=1)$, smeared by a unit (and hence constant) lapse function $N$. Since the operator correspondent of $V$ is rigorously
constructed as an operator in LQG  \cite{rsvol,alvol,ttvol,jurekvol} and since proposals are available for the construction of the Euclidean Hamiltonian constraint operator \cite{qsd,habitat}, it is possible to attempt the construction 
of Thiemann's complexifier in the quantum theory as the commutator between the operators corresponding to $V$ and $H_E(N=1)$. The finite transformation is then mediated in quantum theory by the exponential
of $\frac{1}{i\hbar}$ times the operator correspondent of the complexifier. More precisely, if the integral of the trace of the extrinsic curvature multiplied by a real factor of $\frac{\pi}{2G}$ is denoted by $C_T$, the canonical transformation is generated
by $-iC_T$ and the action of this transformation in quantum theory on any operator ${\hat O}$ is $e^{\frac{\hat C_T}{\hbar}}{\hat O} e^{-\frac{\hat C_T}{\hbar}}$ where ${\hat C}_T$ is the operator correspondent of $C_T$ \cite{ttwick}. 
In particular, the Lorentzian Hamiltonian constraint ${\hat h}_{L}(x)$ is obtained 
from its Euclidean counterpart ${\hat h}_{E}(x)$  as ${\hat h}_L (x)= e^{\frac{\hat C_T}{\hbar}}{\hat h}_E(x) e^{-\frac{\hat C_T}{\hbar}}$ and the physical states, $\Psi_L$, which are annihilated by the {\em dual} action of the Lorentzian constraints can be obtained from their Euclidean counterparts $\Psi_E$
as $\Psi_L= e^{-\frac{\hat C_T}{\hbar}} \Psi_E$. Thus as suggested by Ashtekar \cite{aawick} one may attempt to construct the physical states of the Lorentzian theory within the Hilbert space of the Euclidean theory.

The key issue to be resolved with regard to the above proposal is that of 
the well definedness of the map between the Euclidean and Lorentzian quantum theories. Clearly, the well definedeness or lack thereof hinges on
the properties of the operator ${\hat C}_T$ corresponding to the complexifier and of its exponential $e^{-\frac{\hat C_T}{\hbar}}$.
In this regard, we note the following.
First, as emphasised by Thiemann, the 
complexifier function $C_T$ is not endowed with a particular sign i.e. it can be positive or negative as it varies over phase space. As a result the action of the  exponential operator $e^{-\frac{\hat C_T}{\hbar}}$ is not 
sufficiently under control and its  well definedness  on (putative) physical states of the Euclidean theory is not clear. 
If instead, a complexifier function $C$ could
be found which was {\em positive},  $e^{-\frac{\hat C}{\hbar}}$  would be expected to be  a {\em bounded} operator and hence much better behaved. 

Second, the Euclidean Hamiltonian constraint is constructed in LQG  through a 2 step procedure. In the first step,  a discrete
finite triangulation 
approximant to the constraint is constructed as an operator on the kinematic Hilbert space of LQG  spanned by spin network states. In the second step,  the finite triangulation operator action is evaluated in the  `continuum' limit 
of infinitely fine  triangulation. It is important to note that this limiting action of finite triangulation operators  does not exist in an operator  topology defined by the kinematic Hilbert space norm; a different topology must be used 
\cite{qsd,ttbook,u13mect}.
Since ${\hat C_T}$ is constructed as the commutator between the volume and Hamiltonian constraint operators, it is also defined first at finite triangulation and then in the continuum limit.
Similar procedures must be employed to construct its exponential. The main problem which arises is that adjointness properties  of the finite triangulation operators
on the kinematic Hilbert space do not necessarily survive the continuum limit \cite{habitat} and hence it is difficult 
to exercise control on the adjointness properties of 
continuum limit operators by controlling the adjointness properties of their finite triangulation approximants at the kinematic level.


In this work we address both these points. We address the  first through the construction of a {\em positive} complexifier function $C$ which is differentiable everywhere on phase space except on 
 a set of measure zero
 \footnote{\label{fn1}It is difficult to anticipate  the  repercussions of this lack of  classical differentiability  in  quantum theory.  
 However, since the quantum theory hints at a discrete microstructure, classical configurations are expected to 
 be obtained 
 from quantum states  only after some sort of coarse graining. Moreover, the relevant set of (Liouville) measure zero is characterised by $C$ taking a sharp classical value. In the quantum theory one expects fluctuations
 about sharp classical values. For these reasons a view from  quantum theory of classical configurations  may assign a diminished significance to 
 classical properties of a zero measure set of classical configurations. Hence  it seems reasonable to press on 
 regardless of  classical pathologies on sets of measure zero.}
and  the second by shifting the quantum arena from the kinematic Hilbert space to the diffeomorphism invariant Hilbert space of LQG.
\footnote{Two alternatives to the route advocated in this paper exist to address the adjointness issue: the first  based on the new Hilbert space introduced by Lewandowski and Sahlmann\cite{jurekhanno} and the second on Thiemann's symmetric operator \cite{qsd2}. 
We comment on these in section \ref{sec5}.
We also note that the route advocated in this paper may already have been implicitly suggested in earlier works such as \cite{qsd2}.}


The layout of this paper is as follows. In section \ref{sec2} we construct the new positive complexifier on phase space and discuss its properties. In section \ref{sec3} we argue that it is necessary  to shift the primary quantum arena for 
the analysis of the properties of quantum complexifiers from the kinematic to the diffeomorphism invariant Hilbert space. 
Next, we comment on the construction of the 
operator version of the positive complexifier on the diffeomorphism invariant Hilbert space as well as on the implementation of the Wick rotation proposal with this choice of complexifier. In section \ref{sec4}
we discuss issues related to the existence of the zero measure set of complexifier  non-differentiability.
Section \ref{sec5} contains a summary of our results, a discussion of possible alternatives to tackle the self adjointness issue and speculations on the path ahead.

\section{\label{sec2} A positive complexifier}

\subsection{\label{sec2.1}The real and complex phase space variables: Notation and Review}
We provide a brief review of the Ashtekar Barbero variables and the complex Ashtekar variables mainly to set up notation and define conventions. The reader may consult \cite{fer,aanv} for details.
We use the conventions of Reference \cite{ttwick}.
The real Ashtekar-Barbero phase space variables are an $SU(2)$ connection $A_{a}^i$ and its conjugate unit density weighted electric field $\E^a_i$ with non-vanishing Poisson bracket:
\be
\{A_{a}^i (x), \E^b_j (y)\}= G\delta^a_b \delta^{i}_j\delta (x,y)
\label{aepb}
\ee
where $a$ is a tangent space indice on the Cauchy slice $\Sigma$, $i$ is an $SU(2)$ internal index and $G$ is Newton's constant. The Cauchy slice $\Sigma$ is assumed to be compact without boundary.
We shall raise  and lower  internal indices  by the Kronecker delta $\delta_{ij}$ (in other words $-\delta_{ij}$ is the Cartan-Killing form for $SU(2)$).
Here $\E^a_i$ has the interpretation of a densitized triad field 
so that $\E^a_i\E^{bi}= qq^{ab}$ where $q$ is the determinant of the metric. It is easy to see that $q_{ab}$ can be re-constructed from $\E^a_i$.

The $SU(2)$ connection takes the form:
\be
A_{a}^j = \Gamma_a^j + K_a^j
\label{a=g+k}
\ee
where  $\Gamma^i_a$ is the spin connection compatible with the triad and  $K_a^i$ is related to the 
the extrinsic curvature $K_{ab}$  of the slice $\Sigma$ through:
\be
2K_{ab} = K_{ai}e_b^{i} + K_{bi}e_a^{i}
\ee
where $e_b^i = \frac{q_{ab}\E^b_i}{\sqrt q}$  is the cotriad (and where in this relation $q_{ab}$ and its determinant are constructed from $\E^a_i$).
The Gauss Law, Vector and Hamiltonian constraints of Euclidean gravity expressed in these real phase space variables take the form
\ba
{\cal G}_E &=&{\cal D}_a\E^a_i  \label{glawe}\\
V_{Ea} &=& \E^b_iF_{ab}^{i} \label{ve}\\
h_E &=&  \epsilon^{ijk} \E^a_i\E^b_j F_{abk} \label{hame}
\ea
where $\epsilon^{ijk}$ is the alternating tensor (related to the structure constants of the Lie algebra of  $SU(2)$), ${\cal D}$ is the gauge covariant derivative associated with $A_a^i$ and $F_{ab}^i$ is the curvature of $A_{a}^i$.

Next, we turn to  Lorentzian gravity and the Ashtekar variables.
The Ashtekar momentum variable is,  as in the real case, the real densitised triad $\E^a_i$. The  connection variable is complex and given by  
\be
\Ap_{a}^j = \Gamma_a^j - i K_a^j
\label{a=g-ik}
\ee
where  $i$ in the above equation refers to the square root of $-1$ and should not be confused with the symbol for an $SU(2)$ index (in what follows the context will make it amply clear and there will be no room for
confusion). Here $\Gamma_a^j, K_a^j$ are exactly the same fields as employed in the discussion of the real variables above.

The Gauss Law, Vector and Hamiltonian constraints of Lorentzian gravity expressed in these Ashtekar variables are:
\ba
{\cal \Gp}_L &=&{\cal \Dp}_a\E^a_i  \label{glawp}\\
\Vp_{aL} &=& \E^b_i\Fp_{ab}^{i} \label{vp}\\
\hp_L &=&  \epsilon^{ijk} \E^a_i\E^b_j \Fp_{abk} \label{hamp}
\ea
where    ${\cal \Dp}$ is the gauge covariant derivative associated with $\Ap_a^i$  and    $\Fp_{ab}^i$ is the curvature of $\Ap_{a}^i$.
The  variables $(\Ap_{a}^j , i\E^b_k) $ are canonically related to the real variables so that the only non-vanishing Poisson Bracket is:
\be
\{\Ap_{a}^i (x), \E^b_j (y)\}= -iG\delta^a_b \delta^{i}_j\delta (x,y)
\label{aeppb}
\ee

The connection (\ref{a=g-ik}) is the pull back of the {\em self dual} part of the 4d spin connection \cite{aanv} and on-shell is one of the Sen connections \cite{aanv,sen}.
As noted by Ashtekar \cite{aanv} one could {\em equally} well use a second  connection which is the pull back of the {\em anti-self dual} part of the 4d spin connection (and is on-shell the second Sen connection \cite{aanv,sen}). 
We shall refer
to this `anti self-dual' connection as  $\Am_a^i$ and it is given by 
\be
\Am_{a}^j = \Gamma_a^j + i K_a^j .
\label{a=g+ik}
\ee
In obvious notation, the polynomial constraints (\ref{glawp})-(\ref{hamp}) are equivalent to the set of polynomial constraints:
\ba
{\cal \Gm}_L &=&{\cal \Dm}_a\E^a_i  \label{glawm}\\
\Vm_{aL} &=& \E^b_i\Fm_{ab}^{i} \label{vm}\\
\hm_L &=&  \epsilon^{ijk} \E^a_i\E^b_j \Fm_{abk} \label{hamm}
\ea
with (\ref{aeppb}) replaced by 
\be
\{\Am_{a}^i (x), \E^b_j (y)\}= iG\delta^a_b \delta^{i}_j\delta (x,y)
\label{aempb}
\ee

As remarked in Reference \cite{aabook}, while there is no reason to prefer the self dual variables over the anti-self dual ones in classical theory, one seems to be forced to make 
a choice when one proceeds to a `connection representation' in the quantum theory with the deeper reason for the necessity of such a choice not understood. 
As we shall see below, the employment of a positive  complexifier function to define the Lorentzian theory from the Euclidean one restores the democratic use of the self dual and
anti-self dual variables in quantum theory.

\subsection{\label{sec2.2}Review of Thiemann's complexifier}
We briefly review the essential features of Thiemann's remarkable work. The reader is urged to consult References \cite{ttwick,tttransform} for details.
Thiemann's complexifier  is:
\be
T_+ = \frac{\pi}{2G}\int_{\Sigma} K_a^i\E^a_i .
\label{deftp}
\ee
This function multiplied by a factor of $-i$ generates a complex canonical transformation from the real variables $(A_a^i, E^b_k)$ to the complex canonical pair $(\Ap_a^i, i\E^b_k)$:
\ba
\Ap_a^j &= &   \sum_{n=0}^\infty \frac{ \{A_a^j, (-iT_+)\}_{(n)}}{n!} \label{221}\\
i\E^a_j &= &   \sum_{n=0}^\infty \frac{ \{\E^a_j, (-iT_+)\}_{(n)}}{n!} \label{222}
\ea
where $\{A,B\}_{(n)}$ refers to the $n$th order Poisson bracket $\{...\{\{A, B\},B\}...B\}$ with $B$ appearing $n$ times and with  $\{A,B\}_0$ defined to be equal to $A$.
Equations (\ref{221}) and (\ref{222})  follow from the remarkable facts that $(K_a^i, E^b_k)$ are related to  $(A_a^i, E^b_k)$ through a canonical transformation \cite{aabook} and that $T_+$ Poisson commutes with the spin connection 
$\Gamma_a^i$ \cite{ttwick}.

Upto operator ordering this implies that the operator correspondent ${\hat O}_E$ of any function $O_E (A_a^i, \E^b_k)$  of the real variables is mapped to the operator correspondent ${}^{(+)}{\hat O}_{L}$ of the {\em same} function of the
self dual variables $O_E(\Ap_a^i, i\E^b_k)$ through:
\be
{}^{(+)}{\hat O}_{L} = e^{\frac{\hat T_+}{\hbar}} {\hat O}_E e^{-\frac{\hat T_+}{\hbar}}  
\label{owickp}
\ee
It is then straightforward to see that the classical Euclidean constraints (\ref{glawe})- (\ref{hame}) are mapped, upto overall factors  to their Lorentzian counterparts (\ref{glawp})-(\ref{hamp}) by the canonical transformation 
(\ref{221}),(\ref{222}). The idea is to then {\em define} the Lorentzian constraint operators in quantum theory as images of their Euclidean counterparts using equation (\ref{owickp}).
Following Thiemann \cite{ttwick} and Ashtekar \cite{aawick}, we refer to the canonical transformation (\ref{221})-(\ref{222}) and its quantum counterpart (\ref{owickp}) as a {\em Wick Rotation}.

It then follows that, formally, physical states
of Lorentzian Quantum Gravity  (i.e states which lie in the kernel of the Lorentzian constraints) can be obtained by Wick rotating physical states 
of Euclidean Quantum Gravity via
\be
\Psi_{phys, L} = e^{-\frac{\hat T_+}{\hbar}} \Psi_{phys, E},
\label{statewickp}
\ee
where we have used obvious notation.
The equation follows from the definition of the Lorentzian constraints as Wick rotated images of the Euclidean constraints, together with the assumptions 
that ${\hat T}_+$ is  Hermitian (since $T_+$ is real), that the solutions 
lie in the algebraic dual space to an appropriately chosen dense subspace of the kinematic
Hilbert space of LQG and that the constraints are represented on this dual by their adjoints which act by dual action.

Equation (\ref{statewickp}) is formal because of the lack of  adequate control on the well-definedness of the operators ${\hat T}_+$ and $e^{-\frac{\hat T_+}{\hbar}}$.
As remarked earlier, this situation would improve if we could construct complexifiers of definite signs. We do this in the next section.

\subsection{\label{sec2.3} The new complexifiers}

Our starting point is the observation that replacing the complexifier $T_+$ with its {\em negative} results in a canonical transformation to {\em anti-self dual} variables.
Accordingly we define
\be
T_- = -T_+ = -\frac{\pi}{2G}\int_{\Sigma} K_a^i\E^a_i .
\label{deftm}
\ee

It is then immediate to see that $T_-$ generates the canonical transformation:
\ba
\Am_a^j &= & \sum_{n=0}^\infty \frac{  \{A_a^j, (-iT_-)\}_{(n)} }{n!} \label{231}\\
-i\E^a_j &= & \sum_{n=0}^\infty \frac{ \{   \E^a_j, (-iT_-)\}_{(n)}}{n!} \label{232}
\ea

As noted in section \ref{sec2.1}, the Lorentzian constraints can equally well be obtained by replacing the real variables in the expressions for the Euclidean constraints by the anti-self dual variables.
Hence we may equally well define the Lorentzian quantum constraints and their kernel  as the images of their  Euclidean counterparts
by replacing the `$+$' super- and sub- scripts in (\ref{owickp}) and (\ref{statewickp}) by `$-$'  ones respectively. Here, $T_-$ also suffers from not having a definite sign. 

In order to obtain a complexifier with definite sign we define a new complexifier $T$ as follows:
\ba 
T &= & T_+ \;\;\; {\rm when} \; T_+ >0 \label{t=tp} \\
  &=&  T_-= -T_+  \;\;\; {\rm when} \; T_+ <0  \label{t=tm}\\
  &=& 0   \;\;\; {\rm when} \; T_+ =0  \label{t=0}
\ea
In other words we set 
\be
T := \frac{\pi}{2G} |\int_{\Sigma} K_a^i\E^a_i |
\ee
From (\ref{t=tp}),  (\ref{t=tm}) $T$ generates a canonical transformation  to the self dual Ashtekar variables on those parts of phase space where $T_+$ is positive definite and a canonical 
transformation to the anti-self dual Ashtekar variables on those parts of phase space where $T_+$ is negative definite. On the part of phase space where $T_+$ vanishes,
the complexifier function $T$ is not differentiable.  This part of phase space is of co-dimension 1 and hence of measure zero in accordance with our claim in section \ref{sec1}.

We note that we could equally have chosen -$T$ as a complexifier, or indeed any complexifier of the form $\pm |T_+ -B|$ for any real constant $B$. The latter set of complexifiers are then differentiable
and satisfactory everywhere except on the set defined by $T_+=B$.
In what follows we shall restrict attention to the choice $B=0$.

\section{\label{sec3} Wick Rotation in Quantum Theory} 

In section \ref{sec3.2} we argue that the appropriate arena to define and analyse the properties of the complexifier in quantum theory is the diffeomorphism invariant Hilbert space rather
than the kinematic Hilbert space. For concreteness we phrase our discussion in terms of the Thiemann complexifier $T_+$. In section \ref{sec3.3} we comment on the construction 
of the positive complexifier  $T$  as a positive operator on the diffeomorphism invariant Hilbert space. In section \ref{sec3.4} we assume the existence of such an operator 
and show that its positivity enables the implementation of the `domain changing' strategy suggested by Thiemann \cite{ttwick}, the aim of which is a precise specification of the distributional properties of the
Lorentzian physical states obtained by Wick rotation of Euclidean physical states.
We start in section \ref{sec3.1} with a quick review of our notation  for the  various spaces which are used to house quantum states in LQG.

Before we start, we mention a caveat to our considerations in sections \ref{sec3.2} - \ref{sec3.4}. In these sections we have used properties of operators 
which are known to hold for separable Hilbert spaces. Due to our lack of knowledge, we are not sure if these properties hold for the case of non-separable 
Hilbert spaces considered here. However, even if some of our analysis is questionable in the non-separable context, we believe that it may be possible to
restrict attention to separable subspaces of physical interest and that our analysis would then be applicable to these separable sectors.

\subsection{\label{sec3.1}Review and Notation}

We denote the finite span of spin network states by ${\cal D}$. The space ${\cal D}$ is dense in the kinematic Hilbert space ${\cal H}_{kin}$. The algebraic dual
to any dense set is denoted by a `$*$' superscript so that ${\cal D}^*$  is the algebraic dual space to ${\cal D}$ i.e. ${\cal D}^*$ is the space of complex linear maps on ${\cal D}$.
The diffeomorphism invariant space is constructed by the group averaging of states in ${\cal D}$ \cite{alm2t,aajurekreview,apbook}. The finite span of states, each such state
being the group average of a spin net state, is denoted by ${\cal D}_{diff}$. The completion of  ${\cal D}_{diff}$ in the inner product defined by the group averaging map
yields the diffeomorphism invariant Hilbert space ${\cal H}_{diff}$. The algebraic dual space to ${\cal D}_{diff}$ is denoted by ${\cal D}_{diff}^*$.
Given a representation of a $*$ algebra of operators on the dense domain $D$ of a Hilbert space, the dual action of any operator ${\hat A}$ in this algebra on an element $\Psi$ of the 
algebraic dual $D^*$ is defined to be ${\hat A}\Psi (\phi) = \Psi ({\hat A}^{\dagger} \phi), \;\;\forall \phi \in {\cal D}$ where we have assumed that ${\hat A}, {\hat A}^{\dagger}$
map $D$ to itself.

\subsection{\label{sec3.2}An appropriate arena for the quantum complexifier}

We focus for concreteness on the Thiemann complexifer ${ T}_+$.  As noted by Thiemann \cite{qsd,ttbook}, this complexifier can be expressed as the Poisson bracket between 
the functions  $H_E$  and $V$ 
where 
$V$ is the total volume of space,
\be
V = \int_{\Sigma} \sqrt{q}
\ee
and $H_E$ is the Euclidean Hamiltonian constraint of density one integrated over $\Sigma$ against a unit lapse:
\be
H_E = \int_{\Sigma}\frac{h_E}{\sqrt q}
\ee
where $h_E$ is defined in equation (\ref{hame}). Following Thiemann \cite{qsd,ttbook} and using the conventions of section \ref{sec2} and Reference \cite{ttwick}, we have:
\be
T_+ := \frac{\pi}{2G^2}\{H_E,V\}
\ee
It follows \cite{qsd} that the operator ${\hat T_+}$ 
can be defined upto overall factors as the commutator between the operator correspondents of  ${\hat V}, {\hat H}_E$ of 
$V,H_E$:
\be
{\hat T}_+ := \frac{\pi}{2G^2}\frac{ [{\hat H_E}, {\hat V}] }{i\hbar}
\label{defcomt}
\ee
While the volume operator ${\hat V}$ is well defined \cite{alvol} on the kinematic Hilbert space ${\cal H}_{kin}$ of LQG, the Euclidean Hamiltonian constraint operator ${\hat H}_E$ is not.
Instead \cite{qsd,ttbook}, the operator is first defined at finite triangulation on the kinematic Hilbert space, and then an appropriate continuum limit is taken.
Thus if we were to continue to work in the arena provided by the kinematic Hilbert space and define physical states as states in an appropriate algebraic dual, we would have to 
work first at a finite triangulation characterised by some coarseness parameter $\delta$  with operators   ${\hat H}_{E,\delta}$, ${\hat T_{+, \delta}}$ 
and then take appropriate $\delta\rightarrow 0$ limits in some operator topology such as the `Uniform Rovelli Smolin Operator Topology' (URST) \cite{ttbook}.

Since $H_E,T_+$ are real functions we would like their operator correspondents to be self adjoint. Since ${\hat V}$ is self adjoint \cite{alvol}, from (\ref{defcomt})  a construction of ${\hat H}_E$ as a self adjoint (or 
even symmetric) operator would provide a starting point for the construction of ${\hat T}_+$ as a self adjoint operator. If we could construct ${\hat T}_+$  as a self adjoint operator, its exponential could 
be defined through spectral theory and the properties of the Wick rotation map (\ref{owickp}), (\ref{statewickp}) could then be analysed.
If we adopt the URST view of the continuum limit, then we would like to control the adjointness properties of ${\hat H}_E$ at finite triangulation by choosing its finite triangulation approximants to be 
self adjoint on ${\cal H}_{kin}$. This may be done by choosing any finite triangulation approximant to ${\hat H}_E$ and setting the desired finite triangulation approximant to be half the sum of the 
the chosen  approximant and its adjoint on ${\cal H}_{kin}$. We may then hope that the continuum limit operator is also self adjoint.
Unfortunately, as shown in the beautiful work of Reference \cite{habitat}, the action of the adjoint of the chosen approximant typically {\em vanishes} in the continuum limit so that the 
kinematic self adjointness property at finite triangulation does not survive the continuum limit.

The underlying reason for this trivialization of the action of the kinematic adjoint operators in the continuum limit is as follows. The action of finite triangulation  constraint operators on a spin net state typically creates states with new `offspring' vertices in a $\delta$ vicinity of the `parent' vertex.
Their adjoints when acting on a state check if the state has  a  vertex configuration corresponding to offspring vertices separated by $\delta$ and contract such vertices to yield a parent vertex.
Thus  for a nontrivial action of the adjoint at parameter value $\delta$ on a given state, the state vertex configuration must exactly match an offspring vertex configuration at that precise value of $\delta$. Clearly on a {\em fixed state}
for sufficiently small values of $\delta$ there will be no such configuration so that the continuum limit action of the kinematic adjoint typically vanishes (see \cite{habitat} for a details). 
Preliminary calculations show that key contributions to the kinematic adjoint of the finite triangulation Hamiltonian constraint also vanish in the continuum limit in the context of the toy model of PFT  \cite{pftham}. 
Since the underlying reason is robust, we expect that this situation will persist even if the action of the Euclidean Hamltonian constraint is modified (relative to that considered in 
\cite{qsd,habitat}) so as to incorporate the lessons of \cite{proppft}.

In order to overcome this problem  we propose that the arena for an analysis of the continuum limit operators be changed from the kinematic Hilbert space ${\cal H}_{kin}$ to the diffeomorphism invariant Hilbert space
${\cal H}_{diff}$. Viewed in this way, the continuum limit operators ${\hat H}_E, {\hat T_+}$ are to be seen as operators on ${\cal H}_{diff}$.
The problem is then that 
kinematic self adjointness of finite triangulation approximants to the Euclidean Hamiltonian constraint on ${\cal H}_{kin}$  does not necessarily translate to self adjointness of their continuum limit operator on 
${\cal H}_{diff}$. 
Indeed, in view of the above discussion, we feel that it
 may be   {\em impossible} to construct a self adjoint/symmetric ${\hat H}_E$  operator on ${\cal H}_{diff}$
as the continuum limit of finite triangulation approximants. 
If, as we expect,  this  turns out to be the case  in Euclidean LQG, then a simple way to construct a self adjoint/symmetric ${\hat H}_E$ operator is as follows: 
(i) construct the continuum limit operator, ${\hat H_{E, \delta\rightarrow 0}}$, of suitable finite 
triangulation approximants as an operator on ${\cal H}_{diff}$ (ii) compute the adjoint of this continuum limit operator on ${\cal H}_{diff}$ and (iii) define ${\hat H}_E$ to be half the sum of  (i) and (ii).

In the above discussion we have implicitly assumed that  ${\hat H_{E, \delta\rightarrow 0}}$ can be constructed as an operator on ${\cal H}_{diff}$. Note that this assumption does not follow from the existence of a URST 
continuum limit of ${\hat H_{E, \delta\rightarrow 0}}$. The existence of such a URST limit implies that ${\hat H_{E, \delta\rightarrow 0}}$ maps any diffeomorphism 
invariant distribution which lies in ${\cal D}^*$ to an element of ${\cal D}^*$. However, since $H_E$ is a diffeomorphism invariant function, it is reasonable to require that its operator correspondent 
${\hat H_{E, \delta\rightarrow 0}}$ maps any  diffeomorphism invariant element of ${\cal D}^*$ to another such diffeomorphism invariant element.
Note however, that such elements of ${\cal D}^*$  could in principle lie outside ${\cal H}_{diff}$; for example they could be infinite linear combinations of 
states in ${\cal D}_{diff}$ which are not normalizable in the inner product on ${\cal H}_{diff}$.
Here we make the following simplifying assumption:
\footnote{We shall discuss this assumption further in section \ref{sec5}.}\\

\noindent{\em Assumption A}:\\
\noindent {\em A.1} The continuum limit operator ${\hat H_{E, \delta\rightarrow 0}}$ can be constructed as a densely defined  operator on ${\cal H}_{diff}$.\\
\noindent {\em A.2} ${\cal D}_{diff}$ is a dense domain for ${\hat H_{E, \delta\rightarrow 0}}$.\\
\noindent {\em A.3} ${\hat H_{E, \delta\rightarrow 0}}$  maps ${\cal D}_{diff}$ to itself.\\


One possibility to construct ${\hat T_+}$ as a symmetric operator  on ${\cal H}_{diff}$ is to construct ${\hat H}_E$ as a symmetric operator
through (i)-(iii) above and to
define 
${\hat T_+}$ through
\be
{\hat T_+} := \frac{\pi}{2G^2}\frac{ [{\hat H_E}, {\hat V}] }{i\hbar}
\label{t++}
\ee
This is well defined provided there are no domain issues. More in detail ${\hat V}$ maps ${\cal D}_{diff}$ to itself and from Assumption A above, so does ${\hat H_{E, \delta\rightarrow 0}}$ .
However a domain problem could occur if  ${\hat H_{E, \delta\rightarrow 0}}^{\dagger}$ does not preserve ${\cal D}_{diff}$.
A way to avoid this possible problem  is to first define  the right hand side of (\ref{t++}) using only the contribution (i), 
and define ${\hat T}_+$ as  half of the sum of this right hand side and its adjoint:
\ba
{\hat T_{+,1}} &:=& \frac{\pi}{2G^2}\frac{ [{\hat H_{E,\delta\rightarrow 0}}, {\hat V}] }{i\hbar}\label{t1def}\\
{\hat T_{+}}& :=&
\frac{ {\hat T}_{+,1}  +  {\hat T}_{+,1}^{\dagger}    }{2}
\label{tsdef}
\ea
The existence of  ${\hat T}_+$ as a symmetric operator depends on the  domain of $ {\hat T}_{+,1}^{\dagger}$. If ${\hat T}_+$ is symmetric, one may attempt to find a self adjoint extension and  define its exponential through its spectral 
decomposition.
\subsection{\label{sec3.3} On the existence of ${\hat T}$  as a positive operator}

In this section we comment on the construction of the operator correspondent  ${\hat T}$ of the positive complexifier $T$.
If either of equations (\ref{t++}) or (\ref{tsdef}) define ${\hat T}_+$  as a symmetric operator one may seek to construct a self adjoint extension and define 
${\hat T}$ as the square root of ${\hat T_+}^2$ so constructed.
For this we need to  assume that (a) ${\hat T_{+}}$ can be defined as a densely defined symmetric operator and (b) assume that this symmetric operator admits a self adjoint extension.

A simpler way which may turn out to be of more practical value is as follows. 
Define ${\hat T}_+$ simply as ${\hat T}_{+,1}$ (see equation (\ref{t1def})). 
From Assumption A, ${\hat T}_{+,1}$  maps  ${\cal D}_{diff}$ to itself.
Let us assume  that the domain $D(T_{+,1}^{\dagger})$ of ${\hat T}_{+,1}^{\dagger}$  contains ${\cal D}_{diff}$. 
\footnote{This implies  that ${\hat T_{+}}$ in equation (\ref{tsdef}) can be defined as a symmetric operator on the dense domain ${\cal D}_{diff}$. Hence it is in principle a slightly stronger assumption than (a) 
above where the domain was unspecified. However assumption (b) can be dispensed with. }
Preliminary calculations indicate that  this assumption is satisfied for the appropriate analogs of ${\hat T}_{+,1}$ in PFT \cite{pftham,proppft}.
%

Next note that  for any $\psi, \phi \in {\cal D}_{diff}$,
\be 
(T_+\psi, T_+\phi) = (\psi, T_+^{\dagger} T_+ \phi ) = (T_+^{\dagger}T_+ \psi, \phi)
\label{t21}
\ee
where we have used that ${\hat T}_+\phi, {\hat T}_+\psi \in {\cal D}_{diff} \subset D(T_+^{\dagger})$  Equation (\ref{t21}) implies that ${\hat T}_+^{\dagger} {\hat T}_+$ is a positive symmetric operator on ${\cal D}_{diff}$.
Hence we may construct its  (positive) Friedrich's extension \cite{rs2}.
Finally, we may define ${\hat T}$ to be the positive square root of this Friedrich's extension.

\subsection{\label{sec3.4} Wick Rotation with positive ${\hat T}$}

In this section we assume that ${\hat T}$ can be constructed as 
a densely defined positive self adjoint operator on ${\cal H}_{diff}$. Since under this assumption  $e^{-\frac{\hat T}{\hbar}}$ is a {\em bounded} operator, it may turn out that the Wick rotation
is better defined than with ${\hat T}_+$. More in detail,
any Euclidean physical state $\Psi_{E,phys} \in {\cal D}_{diff}^*$ can be Wick rotated to the Lorentzian physical state $\Psi_{L,phys}$
through:
\ba
e^{-\frac{\hat T}{\hbar}}\Psi_{E, phys} &=& \Psi_{L,phys}  \label{Rhsa}\\
\Rightarrow  \Psi_{L,phys} (\phi) &:=&\Psi_{E, phys} (e^{-\frac{\hat T}{\hbar}} \phi )  \forall \phi \in {\cal D}_{diff} \label{Rhs}.
\ea
Here we have assumed, consistent with the lessons from work on toy  models \cite{pftham,u13anomfree},  that 
the physical states of Euclidean LQG lie in ${\cal D}_{diff}^*$.
If the right hand side of (\ref{Rhs})
is finite,    
$\Psi_{L,phys}$ also resides in ${\cal D}_{diff}^*$.
One may hope that the bounded  operator $e^{-\frac{\hat T}{\hbar}}$ is well behaved enough that this indeed is the case. However this can only be ascertained after the Euclidean theory physical states are constructed so we are
unable to say anything more in this direction.


Instead, in the remainder of this section,  we  show  that it is possible to implement (our interpration of) Thiemann's suggestion in Reference \cite{ttwick} to define
a new dense domain $\tilde{{\cal D}}_{diff} \subset {\cal H}_{diff}$ such that physical states of Lorentzian LQG are precisely characterised as elements of the algebraic dual space  $\tilde{{\cal D}}_{diff}^*$.
In order to do this we shall, counter intuitively, use $-T$ as our complexifier. Accordingly we have that 
\be
e^{\frac{\hat T}{\hbar}}\Psi_{E, phys} =: \Psi_{L,phys}  \label{-t1}
\ee
We define the set 
\be
{\tilde {\cal D}}_{diff} := e^{-\frac{\hat T}{\hbar}} {\cal D}_{diff}
\label{defdtilde}
\ee
We now show that ${\tilde {\cal D}}_{diff}$ is dense in ${\cal H}_{diff}$. First note  that since $e^{-\frac{\hat T}{\hbar}}$  is bounded, it is continuous on ${\cal H}_{diff}$  so that the image of any Cauchy sequence
by $e^{-\frac{\hat T}{\hbar}}$ is Cauchy. Hence we have that 
\be
{     \overline{e^{-\frac{\hat T}{\hbar}} {\cal D}_{diff}}  }  \supseteq e^{-\frac{\hat T}{\hbar}}{\cal H}_{diff}
\label{ethdiff}
\ee
Next, let ${\cal D}_{diff}' \subset {\cal H}_{diff}$ be a dense domain for $e^{\frac{\hat T}{\hbar}}$ so that:
\be
e^\frac{\hat T}{\hbar}: {\cal D}_{diff}' \rightarrow R \subseteq {\cal H}_{diff}
\ee
where $R$ is the image of ${\cal D}_{diff}'$ by $e^\frac{\hat T}{\hbar}$ so that $e^{-\frac{\hat T}{\hbar}} R= {\cal D}_{diff}'$ (recall that since $e^{-\frac{\hat T}{\hbar}}$ is bounded it is defined on all of ${\cal H}_{diff}$).
This in turn implies that:
\be
e^{-\frac{\hat T}{\hbar}}{\cal H}_{diff} \supseteq  {\cal D}_{diff}'.
\ee
Using (\ref{ethdiff}) in the above equation yields:
\be
{     \overline{e^{-\frac{\hat T}{\hbar}} {\cal D}_{diff}}  } \supseteq {\cal D}_{diff}'
\ee

Since  ${     \overline{e^{-\frac{\hat T}{\hbar}} {\cal D}_{diff}}  }$ closed and ${\cal D}_{diff}'$ is dense it follows that ${     \overline{e^{-\frac{\hat T}{\hbar}} {\cal D}_{diff}}  } = {\cal H}_{diff}$. 
Equation (\ref{defdtilde}) then implies 
that ${\tilde {\cal D}}_{diff}$ is dense.

Next, we show that $\Psi_{L,phys}$ in (\ref{Rhsa}) is an element of ${\tilde {\cal D}}^*_{diff}$. To do so, let ${\tilde \phi} \in {\tilde {\cal D}}_{diff}$. From equation (\ref{defdtilde}) we have that 
${\tilde \phi} = e^{-\frac{\hat T}{\hbar}}\phi$ for some $\phi\in {\cal D}_{diff}$. It follows that:
\be
\Psi_{L,phys} ({\tilde \phi} ) = e^{\frac{\hat T}{\hbar}}\Psi_{E, phys}({\tilde \phi} )= \Psi_{E, phys}(e^{\frac{\hat T}{\hbar}}{\tilde \phi}) = \Psi_{E, phys} (\phi)
\ee
Thus $\Psi_{L,phys}$ resides in ${\tilde {\cal D}}^*_{diff}$. 

To summarise: Choosing the complexifier to be $-T$ enables us to define the Wick rotated image of any Euclidean solution as a distribution in the algebraic dual to the new dense set 
${\tilde {\cal D}}_{diff}$. 




\section{\label{sec4} Comments on the zero measure set of complexifier non-differentiability.}

\subsection{\label{sec4.1} Behaviour of the positive complexifier $T$ on the $T=0$ surface}
In the main body of the paper we ignored the lack of differentiability of $T$ on the zero measure, co-dimension 1 surface $T=0$. Here we attempt to quantify this 
lack of differentiability by the formal manipulations which follow below, and to interpret the transformation generated on this surface.

Let $\Theta (x)$ be the step function on the real line;
\ba
\Theta (x) &= & 1 , x>0 \\
           &=&  0,  x=0 \\
&=& -1, x<0 .
\ea
It follows from (\ref{t=tp})-(\ref{t=0}) that 
\be
T= T_+ \Theta (T_+)
\label{ttheta}
\ee
Let $f$ be any smooth function on phase space. It follows that:
\be
\{f,T\} = \{f,T_+\} \frac{d}{dT_+} (T_+\Theta (T_+)) := \{f, T_+\} \alpha (T_+)
\ee
where we have set 
\be
\alpha (T_+) := \frac{d}{dT_+} (T_+\Theta (T_+)) = 2T_+ \delta (T_+) + \Theta (T_+) 
\label{defalpha}
\ee
This
implies
\be
\{f,T\}_{(2)}=   \{f,T_+\}_{(2)} (\alpha (T_+))^2 +  \{f,T_+\}\{ \alpha (T_+), T_+\Theta (T_+)\}
\label{alphatheta}
\ee
If $\alpha , \Theta$ were smooth functions of $T_+$ rather than distributions, the last Poisson bracket in (\ref{alphatheta}) would vanish. 
If we assume that there is a way to regulate their Poisson bracket so that it vanishes, we have that 
\be
\{f,T\}_{(2)}=   \{f,T_+\}_{(2)} (\alpha (T_+))^2 .
\ee
We may compute higher order Poisson brackets in a similar fashion. In doing so we encounter
Poisson brackets between pairs  of distributions each of which depend solely on $T_+$. If we assume that there is a way to regulate such Poisson brackets so that they vanish,
we have that 
\be
\{f, T\}_{(n)}= \{f, T_+\}_{(n)} (\alpha (T_+))^n .
\ee
Finally, from its definition (\ref{defalpha}) it seems plausible that a regularization exists such that we may set $\alpha =0$ at $T_+=0$.
This would mean that $\{f, T\}_{(n)}$ vanishes for all $n\geq 1$. Assuming this is true and  using (\ref{221}), (\ref{222}) with $T_+$ replaced by $T$,  we arrive at the conclusion 
that $T$ generates a canonical transformation to self dual variables when $T_+>0$, to anti-self dual variables when $T_+ <0$ and the identity transformation when $T_+=0$.
Hence, if we are to take the above formal arguments seriously the Wick transformation generated by $T$ maps the Euclidean theory to a pair of `Lorentzian phases'
seperated by a `Euclidean phase' boundary. If  we use the complexifier $|T_+-B|$ (see the last paragraph of section \ref{sec2.3}), it is straightforward to repeat our formal manipulations 
above with the substitution $T_+\rightarrow T_+-B$ and conclude that the Euclidean phase boundary shifts to the codimension 1 surface $T_+=B$. We note here that Euclidean phases do
appear in prior discussions of quantum gravity \cite{hartlehawking,bojowald}.

\subsection{\label{sec4.2} Quantum theory viewpoints on  the zero measure set of complexifier non-differentiability}

For the reasons spelt out in Footnote \ref{fn1}, we believe that it is a useful strategy to pursue the construction  of the positive complexifier as an operator in quantum theory despite its
classical non-differentiability on  an infinite dimensional surface of codimension 1, and hence of measure zero, in phase space.
Nevertheless, one must be open to the possibility that this classical lack of differentiability may give rise to unphysical features in the putative quantum theory especially
because (a large subset of) classical initial  data on this codimension 1 surface qualify as benign from a classical point of view.
 
One way to avoid this particular surface is to use the complexifier $|T_+- B|$ (see the end of section \ref{sec2.3}). This moves the pathologicial surface from $T_+=0$ to $T_+=B$.
One may then treat $|B|$ as a regularization parameter to be taken to infinity. It is instructive to examine the location of the pathological surface for large $|B|$.
To this end,  consider a point $p= ({\tilde E}^a_i, K_b^j)$ in phase space  on  the surface $T_+=B$. Let $V$ be the volume of the spatial slice computed from ${\tilde E}^a_i$. 
Define the spatial average of the trace of the extrinsic curvature at $p$ as $\frac{\int_{\Sigma} {\tilde E}^a_i K_a^i}{V}$.
Then  for  $|B|$ large enough that  $\frac{G|B|}{V}  >> \frac{1}{l_P}$ (here $l_P$ is the Planck length), it follows  that
the spatial average of the trace of the extrinsic curvature at $p$ is trans Planckian. Thus as $|B| \rightarrow \infty$, we may ascribe
any feautures arising from non-differentiability  of $|T_+-B|$ to deep quantum gravity physics.
Indeed, from the considerations of section \ref{sec4.1}, one may perhaps conceive of  the existence  of a trans Planckian Euclidean phase as one of the physical imprints of the $|B| \rightarrow \infty$ limit.

Reverting to the choice $B=0$,
 one may envisage an attempt at combining the Dirac quantization strategy adopted in LQG with some sort of gauge fixing. One popular
gauge choice fixes the trace of the extrinsic curvature to be constant on the spatial slice \cite{york}. If this constant is non-zero
then such gauge fixed configurations lie away from the pathological surface of complexifier non-differentiability. On the other hand if the trace of the extrinsic curvature vanishes, these configurations lie exactly on the
pathological surface. 
In this regard we note the curious fact that  
most Lorentzian spacetimes with topology $\Sigma \times R$, $\Sigma$ either
closed or asymptotically flat, which satisfy the Einstein equations with matter sources obeying the weak energy condition, do not admit  spatial slices with vanishing trace of extrinsic curvature
 due to topological obstructions \cite{donwitt}.

\section{\label{sec5} Summary, Open Issues and Possible Strategies}
\subsection{\label{sec5.1}Summary}
The two main points made in this work are:\\
\noindent (1) It is possible to construct a positive complexifier which ascribes  an equal role to the anti-self dual and self dual Ashtekar variables. This comes at the 
cost of introducing an infinite dimensional surface of non-differentiability. Since the surface is of measure zero, it is still a useful strategy to pursue the use of this complexifier 
in constructing Lorentzian LQG via a Wick rotation of Euclidean LQG as proposed by Thiemann.\\
\noindent (2)  The Thiemann complexifier ${\hat T}_+$  is constructed as the commutator between the Hamiltonian constraint operator  smeared with a unit lapse, ${\hat H}_E$, and the total volume operator ${\hat V}$.
Hence it is  necessary to first construct  ${\hat H}_E$. However ${\hat H}_E$
cannot be constructed directly on the LQG kinematic Hilbert space ${\cal H}_{kin}$. Instead it is
necessary to  construct  ${\hat H}_E$
as a continuum limit of kinematically well defined finite triangulation approximant operators.  Since $H_E$ is real, it is desireable that ${\hat H}_E$ be represented as a self adjoint operator,
While it may seem reasonable to expect that a choice of   approximants which are self adjoint on ${\cal H}_{kin}$  leads to 
a continuum limit operator ${\hat H}_E$  which is self adjoint, the analysis of \cite{habitat}  together with a  study of the PFT toy model suggests that this need not be true.
The reason is that  certain contributions to  the action of kinematically  self adjoint approximants vanish in the continuum limit. 
Since $T_+$ is real and since ${\hat T}_+$ is constructed from ${\hat H}_E$, it is expected that a similar adjointness problem afflicts its construction.

We propose to 
get around this problem of the destruction of kinematic adjointness properties in the continuum limit by changing the arena from ${\cal H}_{kin}$ to the diffeomorphism invariant Hilbert space ${\cal H}_{diff}$.
Since $H_E$ is diffeomorphism invariant, 
we assume that its  continuum limit operator correspondent is well defined on the  diffeomorphism invariant Hilbert space ${\cal H}_{diff}$.
The problem of adjointness now takes the form: the continuum limit of kinematically self adjoint finite triangulation approximants is not necessarily self adjoint on ${\cal H}_{diff}$.
The problem admits a ready solution:  construct the continuum limit operator, compute its adjoint with respect to the  Hilbert space structure of ${\cal H}_{diff}$
and construct the desired  self adjoint operator ${\hat H}_E$ as half the sum of the continuum limit operator and its adjoint. 
Since  $T_+$ is diffeomorphism invariant and since ${\hat T}_+$ 
is constructed from ${\hat H}_E$,  it follows from our proposal that 
the appropriate arena 
to analyse quantum complexifier adjointness properties  should be  ${\cal H}_{diff}$ rather than 
${\cal H}_{kin}$. A similar conclusion holds for the analysis of self adjointness and positivity properties of the operator correspondent of the positive complexifier.
\footnote{
We note here that our proposal to get around the adjointness problem may already have been implicit in earlier works  (for e.g. see second paragraph before Definition 3.1 in \cite{qsd2}).
Indeed, one of the purposes of this work is to argue that  the complexifier ideas of Thiemann \cite{ttwick} and their suggested application by Ashtekar \cite{aawick}, both formulated in the early
days of LQG, 
 when re-examined in the light of developments since then,  provide an eminently viable path to the construction of Lorentzian LQG.}
\\

Underlying our analysis of complexifier operator properties is Assumption A of section \ref{sec3}. 
In this regard note that ${\cal D}_{diff}$ and ${\cal H}_{diff}$ are enormous spaces and presumably contain  many states which are not physically relevant. 
Even if Assumption A is not  valid, the hope is that we may be able to restrict attention to some suitable subspace of states within  ${\cal D}_{diff}$ for which an appropriate
version of Assumption A would be valid. Related to this is our view of LQG itself. We view it as a conservative effort rooted in the continuum which provides glimpses of a 
discrete microstructure. At some stage it may be necessary to use the intuition for this discrete microstructure provided by this conservative effort to make radical jumps which take this microstructure as fundamental with  continuum 
structures being emergent. This may entail an enlargement of the gauge group of diffeomorphisms \cite{zapata,rovelli,mediracpft}. It may be that on the resulting 
replacement for ${\cal H}_{diff}, {\cal D}_{diff}$, some version of Assumption A is valid.

\subsection{\label{sec5.2} Comparision with Alternatives: Open Issues}
Our viewpoint on the contiuum limit definition of  operator correspondents of diffeomorphism invariant functions is that the continuum limit operator is 
defined on ${\cal H}_{diff}$ rather than on ${\cal H}_{kin}$, the latter being the `URST' viewpoint articulated in \cite{ttbook}.  As mentioned in section \ref{sec3},
from \cite{habitat} and from preliminary calculations in the PFT toy model, the URST viewpoint does not allow us to 
circumvent the adjointness problem and we expect that even with a modified Euclidean constraint action of the type in \cite{u13anomfree} this situation may not change. 
It would  be useful to see  if one  could suitably refine/modify  the URST viewpoint in a consistent and practically 
useful way so as to analyse the properties of the quantum complexifier staying within ${\cal H}_{kin}$. 

Absent this,  a change in arena from ${\cal H}_{kin}$ seems to be necessary. One possibility is to work in ${\cal H}_{diff}$. One may do this in the manner advocated here.
Another possibility is to use Thiemann's symmetric constraint \cite{qsd2}. As far as we understand, the regulated symmetric constraint is defined 
on a certain modification of ${\cal H}_{kin,mod}$ of ${\cal H}_{kin}$.
The modification seems to require an extension of the piecewise analytic category of spin network graphs
appropriate to ${\cal H}_{kin}$ \cite{qsd2}
to graphs with not only piecewise analytic edges but also 
certain `marked' edges 
which are $C^{\infty}$. 
To our understanding, the continuum limit is not taken but the dual operator corresponding to  the symmetric constraint smeared with unit lapse  
has a well defined action on the space of diffeomorphism invariant
states ${\cal H}_{diff,mod}$  obtained by group averaging spin net states in 
${\cal H}_{kin,mod}$ with respect to diffeomorphisms
\footnote{To our understanding, the  type of diffeomorphisms employed in \cite{qsd2} are ones which preserve analyticity. The current state of art employs semianalytic diffeomorphisms \cite{lost,ttbook}
which preserve only piecewise analyticity of graphs. 
Nevertheless, we believe that it should be possible to find a way to apply the basic idea of `marking' edges in \cite{qsd2} to the semianalytic context.}, 
and, moreover, this dual operator  is symmetric on ${\cal H}_{diff,mod}$.
Hence, it may be considered as a candidate for ${\hat H}_E$. However it is not clear to us if the action of the symmetric constraint is consistent
with the requirement of a {\em non-trivial} anomaly free representation 
of the Poisson bracket between a pair of (higher density) Hamiltonian constraints in the sense of \cite{u13anomfree}.
Another possibility is to work in the new Hilbert space defined by Lewandowski and Sahlmann \cite{jurekhanno}.  The Lewandowski-Sahlmann Hilbert space, remarkably, supports the continuum limit action
of the Hamiltonian constraint discussed in their work for {\em any} choice of lapse, and hence, in particular, for unit lapse. However,   
it is not clear to us if their construction of this Hilbert space can be generalised to support a constraint  action of the type \cite{u13anomfree}
which changes the vertex set of the state acted upon.  
Clearly, a deeper investigation of the
constructions of References \cite{qsd2,jurekhanno} with regard to the non-trivial anomaly free requirement  would be very useful. 

On the other hand, while our proposal to shift the arena for the implementation of the Wick rotation (\ref{statewickp}) from the kinematic  to  a 
diffeomorphism invariant one  {\em is}  applicable to physical states annihilated by the nontrivial anomaly free 
constraint actions of the type \cite{u13anomfree}, 
the following issue arises.
The complexifier and its exponential are defined now only on  ${\cal D}_{diff}, {\cal H}_{diff}$. On the other hand the constraint ${\hat h}_E$ smeared with an arbitrary lapse does not preserve ${\cal D}_{diff}, {\cal H}_{diff}$.
Hence the relation (\ref{owickp})  with ${\hat O}_E$ set equal to this smeared Hamiltonian constraint is not well defined on the diffeomorphism invariant space ${\cal D}_{diff}$ (nor on its algebraic dual ${\cal D}_{diff}^*$).
As a result, the relation (\ref{statewickp}) which {\em does not} suffer from this issue acquires the status of a {\em proposal} with formal motivation deriving from (\ref{owickp}).

One possible route to the construction of physical states
through (\ref{owickp})- (\ref{statewickp}), using only diffeomorphism invariant arenas, which seems worth exploring, is to use  the Master Constraint formalism \cite{mastercon}. Define the diffeomorphism invariant Euclidean and Lorentzian Master Constraints as
$H_{E,M}:= \int \frac{d^3x}{\sqrt{q}} (\frac{h_E}{\sqrt{q}})^2$, $H_{L,M}:= \int \frac{d^3x}{\sqrt{q}} (\frac{h_L}{\sqrt{q}})^2$. Using $T_+$ (or $T$), the vanishing of the Wick rotated image  of $H_{E,M}$ implies $H_{L,M}=0$ 
(or $H_{L,M}=0$ away from the $T=0$ surface). More precisely the Wick rotated image is proportional to $H_{L,M}$ with the exact proportionality depending on how we choose the roots of $i$ coming from the Wick rotation of 
the $\sqrt{q}$ factors in $H_{E,M}$  by $T_+$
(or the roots of $\pm i$  with  the phase space dependent $\pm$ signs depending on whether these factors are rotated by $T$ at $T_+ >0$ or $T_+ < 0$ ). 
%
Our suggestion is to  set  ${\hat O}_E$ in (\ref{owickp}) to be the  Euclidean Master constraint operator and to identify Lorentzian solutions 
with the kernel of the Wick rotated image ${\hat O}_L$ which then implies  
(\ref{statewickp}). 
The consideration of the  many subtelities involved in the Master Constraint program \cite{mastercon,insideview} and their implications for our suggestion here are beyond the scope of this
work and should be analysed. The key issue is if the structure of ${\hat O}_L$ and its (Wick rotated) solutions suggests a way to construct  the constraints   ${\hat h}_L$
(smeared with arbitrary lapse functions)  in such a way that these solutions lie in the kernel of (the smeared)  ${\hat h}_L$ so constructed.

We end this section with a general remark pertaining to Wick rotated operators.  Dynamically important diffeomorphism invariant operators of the Euclidean theory which arise as counterparts of classically 
real functions such as the Euclidean Master Constraint or real Dirac observables are expected to be represented as self adjoint operators on ${\cal H}_{diff}$.  Their Wick rotated counterparts are
then generically {\em not} self adjoint
and it would seem that their  spectral analysis would be extremely involved.  
In this regard, we note that such Wick rotated images of self adjoint operators seem to resemble a class of well studied
operators known as {\em pseudo Hermitian} operators whose spectral properties, despite their non-Hermiticity,   are under mathematical control: for example  a large class of such operators have
eigen values which are  either real or occur in  complex conjugate pairs\cite{ali1}.
While many of these studies are in the context of quantum mechanical operators with discrete spectra with a slightly restrictive notion of pseudo Hermiticity, there is a vast literature on the subject (see for e.g. \cite{ali2,bender}) which
may prove useful  for a spectral analysis of  Wick rotated operators relevant to Lorentzian LQG.

\subsection{\label{sec5.3} Future Work}
First, the complexifier strategies, outlined in this work  and in Thiemann's earlier works, should be tested  on simpler systems which have a diffeomorphism constraint. 
In this regard 2+1 gravity  provides an excellent testing ground for which a beautiful first analysis already exists \cite{brunojacek} and should be built upon. 
Second, as remarked in section \ref{sec5.2}, a possible connection to the rich literature on pseudo Hermitian operators \cite{ali2} should be explored. 
%
Third, it is imperative to make progress in the construction of Euclidean LQG. We believe the state of the art now permits a serious re-engagement with this problem.

An ever present question is what to do even if we have constructed the Euclidean LQG solution space. Since there are no Dirac observables, how must we interpret these states and how do we
construct the correct inner product on this space? It seems to us that any intepretation rests on a satisfactory interpretation of states in ${\cal D}_{diff}$. If we could construct semiclassical
states in ${\cal D}_{diff}$ and interpret them as quantum Cauchy slices one  could  hope that semiclassical physical states could be built as  linear combinations of such states and that
such physical states could be interpreted as representing quantum spacetimes (in the immensely simpler context of Parameterised Field Theory this is exactly what happens).
A beautiful suggestion for how to construct an inner product on physical states has been made by Thiemann in the mid nineties in \cite{ttlgrav}. It would be good to revisit this suggestion in the light of
developments since then. One would then like to Wick rotate these states and construct Lorentzian states. It is difficult  to anticipate at this stage what the structure of such states could be and 
how to confront the problems of the inner product and interpretation in the Lorentzian theory. Hence we end further speculation  as well as  this paper here.

\section*{\bf Acknowledgements:} We thank Eduardo Villase${\tilde {\rm n}}$or for help with functional analytic issues. We are very grateful to  Fernando Barbero and Thomas Thiemann
for useful conversations and comments on a draft version of this mansucript.



\begin{thebibliography}{999}

\bibitem{adm} R. Arnowitt, S. Deser and C. W. Misner,  {\sl Gravitation: an introduction to current research},  Edited by L. Witten (John Wiley and Sons Inc., New York, London, 1962).

\bibitem{aanv} 
A. Ashtekar,  {\sl Phys.Rev.}{\bf  D36} 1587 (1987). 
 
\bibitem{tedlee} T. Jacobson and L. Smolin,  {\sl Nucl.Phys.}{\bf B299}  295 (1988).

\bibitem{leecarlo} C. Rovelli and L. Smolin, {\sl Nucl.Phys.}{\bf  B331} 80 (1990). 

\bibitem{aabook} A. Ashtekar, {\em Lectures on Non-perturbative Canonical Gravity} (Notes prepared in collaboration with R.S. Tate), World Scientific Singapore  (1991)

\bibitem{aajurekreview}
A. Ashtekar and J. Lewandowski, {\sl Classical and Quantum Gravity} {\bf 21} R53 (2004)


\bibitem{ttbook} T. Thiemann, {\em Modern Canonical Quantum General Relativity}, Cambridge Monographs on Mathematical Physics. Cambridge University
  Press (2007)



\bibitem{gpbook} R. Gambini and J. Pullin, {\sl A First Course in Loop Quantum Gravity},
Oxford University Press (2011)


\bibitem{apbook} {\em Loop Quantum Gravity: The First 30 Years} in {\em 100 Years of General Relativity: Volume 4}, 
Edited by Abhay Ashtekar and Jorge Pullin, World Scientific (2017)


\bibitem{fer} Barbero, J.F., {\sl Phys. Rev.}{\bf D51}  5507 (1995).
\bibitem{qsd}
T. Thiemann, {\em Classical and Quantum Gravity}, {\bf 15} 839  (1998).

\bibitem{mastercon} T. Thiemann,  {\sl Class.Quant.Grav.}{\bf  23} 2211 (2006)


\bibitem{jurekR} E. Alesci, M. Assanioussi and J. Lewandowski, {\sl Phys.Rev.}{\bf  D89} 124017 (2014).

\bibitem{sam} J. Samuel, {\sl Class.Quant.Grav.}{\bf  17} L141 (2000).

\bibitem{ttwick} T. Thiemann, {\sl  Class.Quant.Grav.}{\bf 13} 1383  (1996).

\bibitem{tttransform} T. Thiemann, {\sl Acta Cosmologica}{\bf 21} 145 (1996). 

\bibitem{ttlgrav} T. Thiemann, {\sl Phys.Lett.}{\bf B380} 257 (1996).

\bibitem{rsvol}  C. Rovelli and L. Smolin, {\sl Nucl.Phys.}{\bf  B442} 593 (1995), Erratum: {\sl Nucl.Phys.}{\bf B456} 753  (1995).

\bibitem{alvol} A. Ashtekar and J.Lewandowski, {\sl Adv.Theor.Math.Phys.}{\bf  1} 388  (1998).

\bibitem{jurekvol} J. Lewandowski,  {\sl Class.Quant.Grav.}{\bf  14} 71  (1997).


\bibitem{ttvol} K. Giesel and T. Thiemann,
{\sl Class.Quant.Grav.}{\bf  23} 5667 (2006); {\em ibid}  {\sl Class.Quant.Grav.}{\bf  23} 5693 (2006).



\bibitem{habitat} J. Lewandowski and D. Marolf, {\sl Int.J.Mod.Phys}{\bf D7} 299 (1998).

\bibitem{aawick} A. Ashtekar,  {\sl Phys.Rev.}{\bf  D53} 2865 (1996).



\bibitem{jurekhanno} J. Lewandowski and H. Sahlmann,  {\sl Phys.Rev.}{\bf  D91} 044022 (2015). 

\bibitem{qsd2} T. Thiemann, {\sl Class.Quant.Grav.}{\bf  15} 875  (1998).


\bibitem{u13mect} C. Tomlin and M. Varadarajan,  {\sl Phys.Rev.}{\bf D87} 044039 (2013).

\bibitem{sen} A. Sen, {\sl J. Math. Phys.}{\bf 22} 1718 (1981); {\sl Phys. Lett.}{\bf  119B} 89
(1982).

\bibitem{alm2t}  A. Ashtekar, J.Lewandowski, D. Marolf, J. Mour${\tilde {\rm a}}$o and T. Thiemann, {\sl J.Math.Phys.}{\bf 36} 6456 (1995).

\bibitem{pftham} A. Laddha and M. Varadarajan, {\sl Phys.Rev.}{\bf D83} 025019 (2011).

\bibitem{proppft}  M.  Varadarajan 
{\sl Class.Quant.Grav.} {\bf 34} 015012 (2017)

\bibitem{u13anomfree} M. Varadarajan, {\sl Phys.Rev.}{\bf D97} 106007 (2018).

%


\bibitem{rs2} M. Reed and B. Simon, {\sl Methods of Modern Mathematical Physics, Vol. 2} (Academic Press, 1975).

\bibitem{lost}J. Lewandowski, A. Okolow, H. Sahlmann and T. Thiemann,  {\sl Commun.Math.Phys.}{\bf 267} 703 (2006). 

\bibitem{zapata} J.A. Zapata,  {\sl Gen.Rel.Grav.}{\bf  30} 1229  (1998); {\em ibid}  {\sl J.Math.Phys.}{\bf 38} 5663  (1997). 

\bibitem{rovelli} W. Fairbairn and C. Rovelli, {\sl J.Math.Phys.}{\bf 45} 2802  (2004).

\bibitem{mediracpft} M. Varadarajan, {\sl Phys.Rev.}{\bf D75} 044018 (2007). 

\bibitem{hartlehawking}
J. Hartle, S. Hawking, {\sl Phys.Rev.}{\bf D28} 2960 (1983).


\bibitem{bojowald}  M. Bojowald and  G. Paily, {\sl Phys.Rev.}{\bf  D86} 104018 (2012).

\bibitem{york} J. W. York, Jr, {\sl Phys.Rev.Lett.}{\bf  28} 1082 (1972); 	
Y. Choquet-Bruhat and  J. W. York, Jr
 in  {\sl General Relativity and Gravitation: An Einstein Centenary Survey}, edited by A. Held (Plenum, New York, 1980), Vol. 1, p. 99. 

 
\bibitem{donwitt} D.M. Witt, {\sl Phys.Rev.Lett.}{\bf  57} 1386  (1986); {\em ibid}, e-Print: arXiv:0908.3205.

\bibitem{insideview} T. Thiemann,  {\sl Lect.Notes Phys.} {\bf 721} 185 (2007)  

\bibitem{bender} C. Bender,{\sl Rept.Prog.Phys.}{\bf  70} 947 (2007); 	
P. Dorey, C. Dunning, R. Tateo in {\em Statistical Field Theories, NATO Science Series II, Vol. 73} edited by A. Cappelli and G. Mussardo,
Springer, Netherlands (2002), 
also available as e-Print: hep-th/0201108.

\bibitem{ali1} A. Mostafazadeh, {\sl J. Math. Phys.}{\bf  43} 205 (2002).

\bibitem{ali2} A. Mostafazadeh, {\sl Int.J.Geom.Meth.Mod.Phys.}{\bf 7} 1191 (2010).

\bibitem{brunojacek} 	
B. Hartmann and J. Wisniewski,   {\sl Class.Quant.Grav.}{\bf  21} 697 (2004).
 

\end{thebibliography}
\end{document}